\documentclass{appolb}
\usepackage{graphicx}
\usepackage{amsmath}
\usepackage{amssymb}
\usepackage[english]{babel}
\usepackage{url}
\usepackage[colorlinks=true,linktocpage=true,linkcolor=blue,citecolor=blue,allcolors=blue]{hyperref}

\newcommand{\bs}{\boldsymbol}

\newcommand{\eq}[1]{\begin{align} #1 \end{align}}
\newcommand{\mean}[1]{\langle #1 \rangle}

\begin{document}
\title{Exploring the QCD phase diagram with fluctuations %
\thanks{Presented at the 60$\mbox{}^{\rm th}$ anniversary of the  Cracow School of Theoretical Physics}%
}
\author{Volker Koch, Volodymyr Vovchenko
\address{Nuclear Science Division, \\ Lawrence Berkeley National Laboratory, \\ 1 Cyclotron Road,
  Berkeley, CA 94720, USA }
}

\maketitle
\begin{abstract}
We report on recent progress concerning theoretical description of event-by-event fluctuations in heavy-ion collisions.
Specifically we discuss a new Cooper-Frye particlization routine -- the subensemble sampler -- which is designed to incorporate effects of global conservation laws, thermal smearing and resonance decays on fluctuation measurements in various rapidity acceptances.
First applications of the method to heavy-ion collisions at the LHC energies are presented and further necessary steps to analyze fluctuations from RHIC beam energy scan are outlined.
\end{abstract}
%\PACS{PACS numbers come here}
  
\section{Introduction}

Fluctuations of conserved charges represent one of the central observables for probing the phase structure of QCD.
The corresponding measurements are in the focus of the experimental search for the QCD critical point at RHIC beam energy scan~\cite{Bzdak:2019pkr,Adam:2020unf}, being expected to exhibit strong critical behavior in its vicinity~\cite{Stephanov:1998dy,Stephanov:1999zu}.
Fluctuations of conserved charges are also
studied in heavy ion experiments at the highest RHIC and LHC energies with the goal to
identify experimentally the remnants of the chiral criticality at vanishing chemical
potentials~\cite{Friman:2011pf,Citron:2018lsq}.

Theoretical calculations of fluctuations are typically performed in the grand-canonical ensemble (GCE), where cumulants of the conserved charge distribution correspond to the susceptibilities -- the derivatives of the grand potential with respect to the chemical potentials.
At zero chemical potentials the QCD susceptibilities are accessible from first principles via lattice QCD simulations~\cite{Bazavov:2017dus,Borsanyi:2018grb}, whereas at finite densities they can be treated with various effective QCD approaches~\cite{Isserstedt:2019pgx,Fu:2019hdw}.
An important question is how to relate the theoretical calculations with experimental measurements.
In the GCE the system can exchange charges with a reservoir, thus the charges are conserved only on average.
In heavy-ion experiments, on the other hand, the charges are globally conserved. It is thus essential to establish how these susceptibilities are related to experimental
measurements~\cite{Bleicher:2000ek,Schuster:2009jv,Bzdak:2012an,Braun-Munzinger:2016yjz,Pruneau:2019baa}. 
A subensemble acceptance method~(SAM) has been developed recently~\cite{Vovchenko:2020tsr,Vovchenko:2020gne}, which allows to correct the grand-canonical cumulants to account for global conservation of (multiple) conserved charges, and for any equation of state, such as that of QCD.

Another important issue is the difference between  coordinate space, where the vast majority of theories operate, and the momentum space, where experimental measurements are performed.
Experimental cuts in the momentum space may correspond to cuts in the coordinate space if there is a strong correlation between the momenta and coordinates of the particles. This is situation for instance in case of longitudinal Bjorken flow, approximately realized at the highest collision energies, where one can associate the kinematic rapidity $Y$ of a particle with its space-time rapidity $\eta_s$ at freeze-out.
Even in this case, however, a smearing of order $\Delta Y_{\rm th} \sim 1$~\cite{Ling:2015yau} due to random thermal motion is present.
Additional smearing is generated by the decays of resonances after the freeze-out.
It is thus crucial to control these effects well for any reliable physics interpretation of experimental measurements.

Recently, a generalized Cooper-Frye particlization routine has been proposed~\cite{Vovchenko:2020kwg}, which allows one to incorporate the aforementioned effects. 
The routine, called \emph{subensemble sampler}, samples the equation of state of an interacting hadron resonance gas that can, for instance, be matched to reproduce the lattice QCD susceptibilities at freeze-out.
Here we discuss the main ideas behind this sampler, the first results obtained for event-by-event fluctuations at the LHC, and the necessary steps for future applications to the measurements in RHIC beam energy scan programme.

\section{Subensemble sampler}

Consider the particlization stage of heavy-ion collisions at the end of hydrodynamic evolution.
This stage is characterized by a hypersurface $\sigma(x)$, where $x$ is the space-time coordinate.
The QCD fluid is transformed into an expanding gas of hadrons and resonances.
The momentum distributions of the all the hadron species at particlization are determined by the famous Cooper-Frye formula, which for hadron species $i$ reads
\eq{\label{eq:CF}
\omega_p \frac{d N_i}{d^3 p} = \int_{\sigma(x)} d \sigma_\mu (x) \, p^\mu \, f_i[u^\mu(x) p_\mu;T(x),\mu_i(x)].
}
Here $T(x)$, $\mu_i(x)$, $u^\mu(x)$, and $d \sigma_\mu(x)$ correspond to the space-time distributions of the temperature, chemical potential, flow velocity, and the hypersurface normal elements, respectively.
$f_j$ is the distribution function in the local rest frame.
We take it in the following form
\eq{
f_i[u^\mu p_\mu;T(x),\mu_i(x)] = \frac{d_i \, \lambda_i(x)}{(2\pi)^3} \, \exp\left[ \frac{\mu_i(x) - u^\mu(x) p_\mu}{T(x)} \right].
}
Here we neglected the viscous corrections. 
The factor $\lambda_i(x)$~($=1$ in the ideal gas limit) incorporates possible deviations of the hadronic equation of state from the ideal gas limit. 
The presence of this factor is one new element of our procedure relative to the standard routine.
In general $\lambda_i$ can be a function of both the space-time coordinate $x$ and the momentum $p$.
Presently we restrict our considerations to the case where $\lambda_i$ depends on $x$ only.

The Cooper-Frye formula~\eqref{eq:CF} defines the momentum distributions of all the hadrons emerging from the hydrodynamic evolution.
This equation, however, carries no information regarding the fluctuations in the event-by-event distribution of the various hadron numbers.
Usually, the yields are sampled in each fluid element from a Poisson distribution.
Because the Poisson distribution is additive, this means that the yields of all hadron species in the full space follow the Poisson distribution as well. 
Most hydro simulations use this type of sampling~\cite{Kisiel:2005hn,Shen:2014vra,Karpenko:2015xea,Bernhard:2018hnz}.
Such a multiplicity distribution, however, is valid only for an ideal Maxwell-Boltzmann hadron resonance gas~(HRG) in the grand-canonical ensemble.
The key new feature of our subensemble sampler is the ability to simultaneously incorporate in the sampling procedure (i) exact global conservation of charges and (ii) interactions between hadrons.

Let us consider for simplicity the case of a single conserved charge $B$.
For the more general case of multiple conserved charges see~\cite{Vovchenko:2020kwg}.
In the subensemble sampler we partition the particlization hypersurface $\sigma$ into contiguous subvolumes.
Let the index $j$ enumerate the subvolumes.
The subvolume $V_j$ then reads $V_j = \int_{x \in \sigma_j} d \sigma_\mu (x) \, u^\mu(x)$.
We further assume that each subvolume is (i) characterized by constant values of the thermal parameters $T$ and $\mu_B$ and (ii) is sufficiently large compared to the correlation length $\xi$, i.e. $V_j \gg \xi^3$.
The latter assumption implies that interactions between particles from different subvolumes can be neglected.
The total partition function can then be written as
\eq{\label{eq:Zce}
Z_{\rm tot}^{\rm ce} & ~ \sim ~ \prod_j \, \sum_{B_j} \, e^{\mu_{B,j}  B_j} \, Z^{\rm iHRG} (T_j,V_j,B_j)\, \times \delta \left(B_{\rm tot} - \sum_k B_k \right)~.
}
Here $Z^{\rm iHRG} (T_j,V_j,B_j)$ is the canonical partition function of an interacting HRG~(iHRG).
Equation~\eqref{eq:Zce} represents a sum over all possible values $B_j$ of the baryon number in each of the subvolumes.
The Kronecker delta ensures that the total baryon number is globally conserved.
One can rewrite Eq.~\eqref{eq:Zce} in terms of the sum over all possible hadron numbers in each of the subvolumes:
\eq{
\label{eq:PNjoint}
P \left(\left\{ \hat{N}_{j} \right\} \right) & = \prod_j \,  P^{\rm iHRG}(\hat{N}_{j}; T_j,V_j,\bs \mu_j) \times \delta \left(B_{\rm tot} - \sum_k B_k \right)~, \\
\label{eq:Qhrg}
B_k & = \sum_{i=1}^f \,  b_i \, N_{k,i}~.
}
Here $\hat{N}_{j} = \{N_{j,i}\}_{i=1}^f$ are hadron numbers in subvolume $j$ and $b_i$ is the baryon charge of hadron species $i$. $P^{\rm iHRG}$ is the probability distribution function of hadron numbers in an interacting HRG.
Each term in Eq.~\eqref{eq:PNjoint} defines the joint probability distribution of all hadrons numbers in all the subvolumes.

The subensemble sampler routine thus consists of the following steps:
\begin{enumerate}
    \item The hadron multiplicities in all the subvolumes are sampled from Eq.~\eqref{eq:PNjoint}. This is achieved by first sampling the grand-canonical multiplicities independently for each subvolume and then applying a rejection sampling step to ensure exact global conservation of baryon number.
    It is assumed that the procedure to sample the grand-canonical multiplicity distribution of the interacting HRG model under consideration is known.
    \item The momenta for each of the hadrons is sampled via the Cooper-Frye formula~\eqref{eq:CF} applied independently for each hadron from each of the subvolumes.
\end{enumerate}

\section{Fluctuations at the LHC energies}

The subensemble sampler has first been applied to event-by-event fluctuations in heavy-ion collisions at the LHC energies in Ref.~\cite{Vovchenko:2020kwg}. 
Here we review the main results that were obtained.

Specifically, we discuss 0-5\% central Pb-Pb collisions at $\sqrt{s_{\rm NN}} = 2.76$~TeV.
As discussed in Ref.~\cite{Vovchenko:2020kwg} the particlization hypersurface in these collisions can be approximated by a longitudinally boost-invariant blast-wave surface covering 9.6 units of space-time rapidity $\eta_s$.
The thermal parameters are uniform and correspond to the freeze-out temperature of $T = 160$~MeV and vanishing chemical potentials.
With this choice the model accurately reproduces the bulk observables in a range $|Y| \lesssim 2$ around midrapidity as well as the total charged multiplicity in full space. 
The latter feature is important to properly account for the effects of global conservation. 
The blast-wave model parameters are taken from Ref.~\cite{Mazeliauskas:2019ifr} ensuring that the $p_T$ distribution of protons at midrapidity is reproduced accurately.
As we are mainly interested in the rapidity dependence of various cumulants, the partition of the particlization hypersurface is performed along the longitudinal space-time rapidity axis. The axis is split into 96 slices~(subvolumes), each covering 0.1 units of rapidity.

The choice of interacting HRG model is constrained by requiring it to agree with lattice QCD data on cumulants of conserved charges that are being studied. For that purpose we take an HRG model with repulsive excluded volume interactions in the baryonic sector, first formulated in Ref.~\cite{Vovchenko:2016rkn}.
As shown in~\cite{Vovchenko:2017xad}, with the excluded volume parameter value of $b = 1$~fm$^3$ this model describes quantitatively both the net-baryon susceptibilities at $\mu_B = 0$ as well as the Fourier coefficients of net baryon density at imaginary $\mu_B$.
The details of the grand-canonical multiplicity sampling within this model are described in Refs.~\cite{Vovchenko:2020kwg,Vovchenko:2018cnf}.
The entire sampling procedure is implemented within an extended version of the open source package Thermal-FIST~\cite{Vovchenko:2019pjl}.

\begin{figure*}[t]
  \centering
  \includegraphics[width=.65\textwidth]{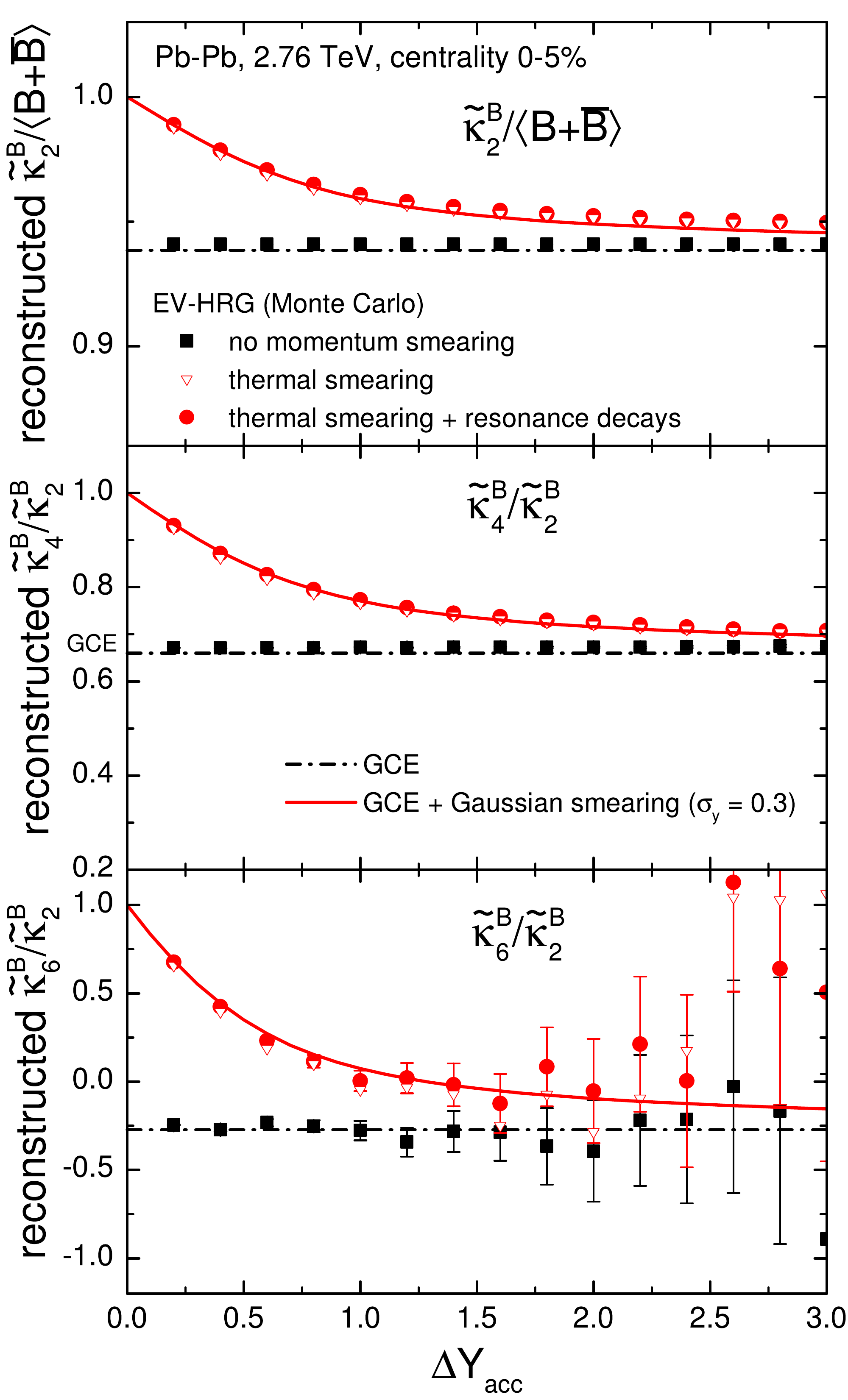}
  \caption{
  Rapidity acceptance dependence of cumulant ratios $\tilde{\kappa}_2^B/\mean{B + \bar{B}}$ (top), $\tilde{\kappa}_4^B/\tilde{\kappa}_2^B$ (middle), and $\tilde{\kappa}_6^B/\tilde{\kappa}_2^B$ (bottom) of net baryon distribution in 0-5\% central Pb-Pb collisions at the LHC in the EV-HRG model and corrected for global baryon conservation via the SAM.
  The symbols depict the results of the Monte Carlo event generator, the full black squares correspond to neglecting the momentum smearing, the open red triangles include the thermal smearing at particlization, and the full red circles incorporate the smearing due to both the thermal motion and resonance decays.
  The solid red lines correspond to a Gaussian rapidity smearing model.
  }
  \label{fig:chiLHC}
\end{figure*}

First we focus on the rapidity dependence of net baryon cumulants.
We analyze the following three ratios:
$$
\frac{\kappa_2[B-\bar{B}]}{\mean{B + \bar{B}}}, \qquad \frac{\kappa_4[B-\bar{B}]}{\kappa_2[B-\bar{B}]}, \qquad \frac{\kappa_6[B-\bar{B}]}{\kappa_2[B-\bar{B}]}.
$$

These ratios are calculated as a function of rapidity by sampling $10^{10}$ events.
The Monte Carlo results naturally incorporate the effect of baryon number conservation on these ratios.
These effects can be corrected for analytically in the framework of the sub-ensemble acceptance method (SAM) of Ref.~\cite{Vovchenko:2020tsr}.
The SAM expresses these cumulant ratios in terms of the corresponding ratios without baryon number conservation effect. 
At the LHC energies, where all odd order cumulants of net baryon number vanish, the SAM expressions read:
\eq{\label{eq:SAMc2}
\left(\frac{\kappa_2[B-\bar{B}]}{\mean{B + \bar{B}}}\right)_{\rm LHC} & = (1 - \alpha) \, \frac{\tilde{\kappa}_2[B-\bar{B}]}{\mean{B + \bar{B}}}, \\
\label{eq:SAMc4}
\left(\frac{\kappa_4[B-\bar{B}]}{\kappa_2[B-\bar{B}]}\right)_{\rm LHC} & = (1-3\alpha \beta) \, \frac{\tilde{\kappa}_4[B-\bar{B}]}{\tilde{\kappa}_2[B-\bar{B}]}, \\
\label{eq:SAMc6}
\left(\frac{\kappa_6[B-\bar{B}]}{\kappa_2[B-\bar{B}]}\right)_{\rm LHC} & =
\left[1-5\alpha \beta (1 - \alpha \beta ) \right] \frac{\tilde{\kappa}_6[B-\bar{B}]}{\tilde{\kappa}_2[B-\bar{B}]} \nonumber \\
& \quad - 10 \alpha (1-2\alpha)^2 \beta \left( \frac{\tilde{\kappa}_4[B-\bar{B}]}{\tilde{\kappa}_2[B-\bar{B}]}\right)^2~.
}
Here $\beta \equiv 1 - \alpha$ and the tilde corresponds to net baryon cumulants without the effect of baryon number conservation.
The parameter $\alpha$ corresponds to the fraction of the total volume covered by the acceptance.
For a $p_T$-integrated acceptance in a range $\Delta Y_{\rm acc}$ around midrapidity in 2.76~TeV Pb-Pb collisions one can approximate this parameter as $\alpha = \Delta Y_{\rm acc} / 9.6$~\cite{Vovchenko:2020kwg}.

Figure~\ref{fig:chiLHC} depicts the $\Delta Y_{\rm acc}$ dependence of the three cumulant ratios where the effect of global baryon conservation was accounted for in accordance with Eqs.~(\ref{eq:SAMc2}-\ref{eq:SAMc6}). 
Here $\tilde{\kappa}_n^B \equiv \tilde{\kappa}_n[B-\bar{B}]$.
The black symbols depict the Monte Carlo results without the effect of momentum smearing, i.e. where the kinematic rapidity of each particle coincides with its space-time rapidity.
In this case the SAM-corrected Monte Carlo results are independent of $\Delta Y_{\rm acc}$ and agree with the corresponding ratios of the grand-canonical susceptibilities, shown in Fig.~\ref{fig:chiLHC} by dash-dotted horizontal lines.

Calculations that incorporate the effect of thermal smearing and resonance decays are depicted by the red symbols.
The effect of resonance decays is found to be largely negligible.
The thermal smearing ``poissonizes'' the cumulants in small acceptances, where the cumulant ratios approach unity in the limit $\Delta Y_{\rm acc} \to 0$.
The deviations of the cumulant ratios from the grand-canonical limit are significant for $\Delta Y_{\rm acc} \lesssim 1$ whereas at larger acceptances they are subleading. 
The magnitude of the effect is similar for all three cumulant ratios considered.
We find that the Monte Carlo results can be accurately described analytically by assuming that the rapidity $y$ of each particle smeared around its space-time rapidity $\eta_s$ by a Gaussian with a width $\sigma_y = 0.3$, the corresponding results are depicted by the red lines in Fig.~\ref{fig:chiLHC}.

The presented results illustrate that a reliable interpretation of the experimental data will require control over the effects of baryon conservation and thermal smearing. 
In particular, as follows from Fig.~\ref{fig:chiLHC}, a measurement of a negative hyperkurtosis $\kappa_6[B-\bar{B}]/\kappa_2[B-\bar{B}]$, corrected for global baryon conservation via the SAM, would indicate a negative grand-canonical hyperkurtosis $\chi_6^B / \chi_2^B$, which could be interpreted as an experimental signature of the chiral QCD crossover transition.
It should also be noted that the obtained results do not incorporate the effect of volume fluctuations~~\cite{Gorenstein:2011vq}.
This effect should be minimized via an appropriate centrality selection and/or corrected for e.g. using the formalism of Ref.~\cite{Skokov:2012ds}.

The experiments do not measure baryons directly but instead use protons as a proxy.
It is natural to expect net protons to carry at least some information about net baryon fluctuations.
In fact, as shown by Kitazawa and Asakawa~\cite{Kitazawa:2011wh,Kitazawa:2012at}, under the assumption of isospin randomization at the late stages of heavy-ion collisions, one can reconstruct the cumulants of net baryon distribution from the measured factorial moments of proton and antiproton distributions.
The subensemble sampler can be used to elaborate on the similarities and differences between net proton and baryon number cumulants.

\begin{figure*}[t]
  \centering
  \includegraphics[width=.65\textwidth]{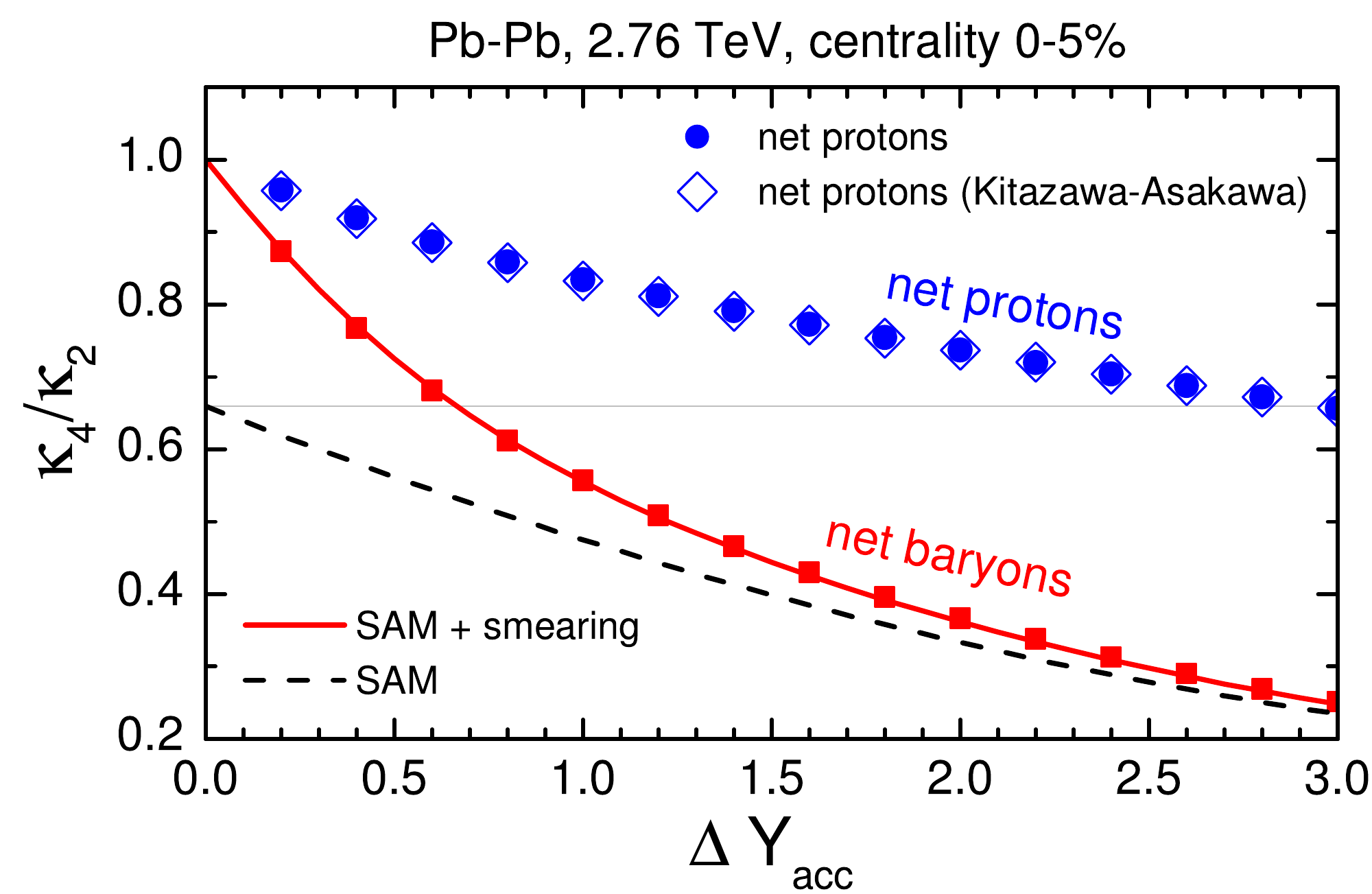}
  \caption{
  Rapidity acceptance dependence of net baryon~(red squares) and net proton~(blue circles) kurtosis $\kappa_4/\kappa_2$ in 0-5\% central 2.76 TeV Pb-Pb collisions at the LHC.
  The open blue diamonds correspond to net proton cumulants evaluated from net baryon cumulants using a binomial folding method of Kitazawa and Asakawa~\cite{Kitazawa:2011wh,Kitazawa:2012at}.
  The analytical predictions of the SAM framework with and without Gaussian rapidity smearing are depicted by solid red and dashed black lines, respectively.
  The thin gray line corresponds to the value in the grand canonical ensemble consistent with lattice QCD.
  Adapted from Ref.~\cite{Vovchenko:2020kwg}.
  }
  \label{fig:netbaryonproton}
\end{figure*}

Figure~\ref{fig:netbaryonproton} depicts the rapidity acceptance dependence of net baryon~(black squares) and net proton~(blue symbols) kurtosis $\kappa_4/\kappa_2$ resulting from Monte Carlo sampling.
The results shown include the effect of global baryon conservation and reveal large differences between net proton and net baryon cumulant ratios.
Net proton cumulant ratios are considerably closer to the Skellam baseline of unity.
This can be understood in the following way.
By taking only a subset of baryons -- the protons -- one dilutes the total signal due to baryon correlations.
This leads to a smaller deviation of cumulants from Poisson statistics -- the limiting case of vanishing correlations.

The results shown in Fig.~\ref{fig:netbaryonproton} clearly indicate that direct comparisons between the grand-canonical cumulants and the measured cumulants of net baryon or net proton distribution are unjustified.
Let us for example take fluctuations within one unit of rapidity.
The predicted kurtosis of net baryon number in $\Delta Y_{\rm acc} = 1$ is affected by thermal smearing and global baryon conservation, resulting in a value $\kappa_4^B/\kappa_2^B \simeq 0.56$.
The value of net proton kurtosis is $\kappa_4^p/\kappa_2^p \simeq 0.83 \neq \kappa_4^B/\kappa_2^B$, considerably different from the net baryon kurtosis.
Both the net proton and net baryon kurtosis differ notably from the grand-canonical net baryon kurtosis of $\chi_4^B/\chi_2^B \simeq 0.67$.

We do observe that the method of Kitazawa and Asakawa~\cite{Kitazawa:2011wh,Kitazawa:2012at} can accurately relate net baryon and proton cumulants between each other.
This confirms that cumulants of net baryon distribution can be recovered from factorial moments of net proton distribution via a binomial unfolding with probability $q$, where $q$ is the ratio between the mean numbers of protons and baryons.
Only the experiment can perform this unfolding model-independently because the factorial moments of baryon (proton) distribution cannot be computed in lattice QCD.

\begin{figure}[t]
  \centering
  \includegraphics[width=.49\textwidth]{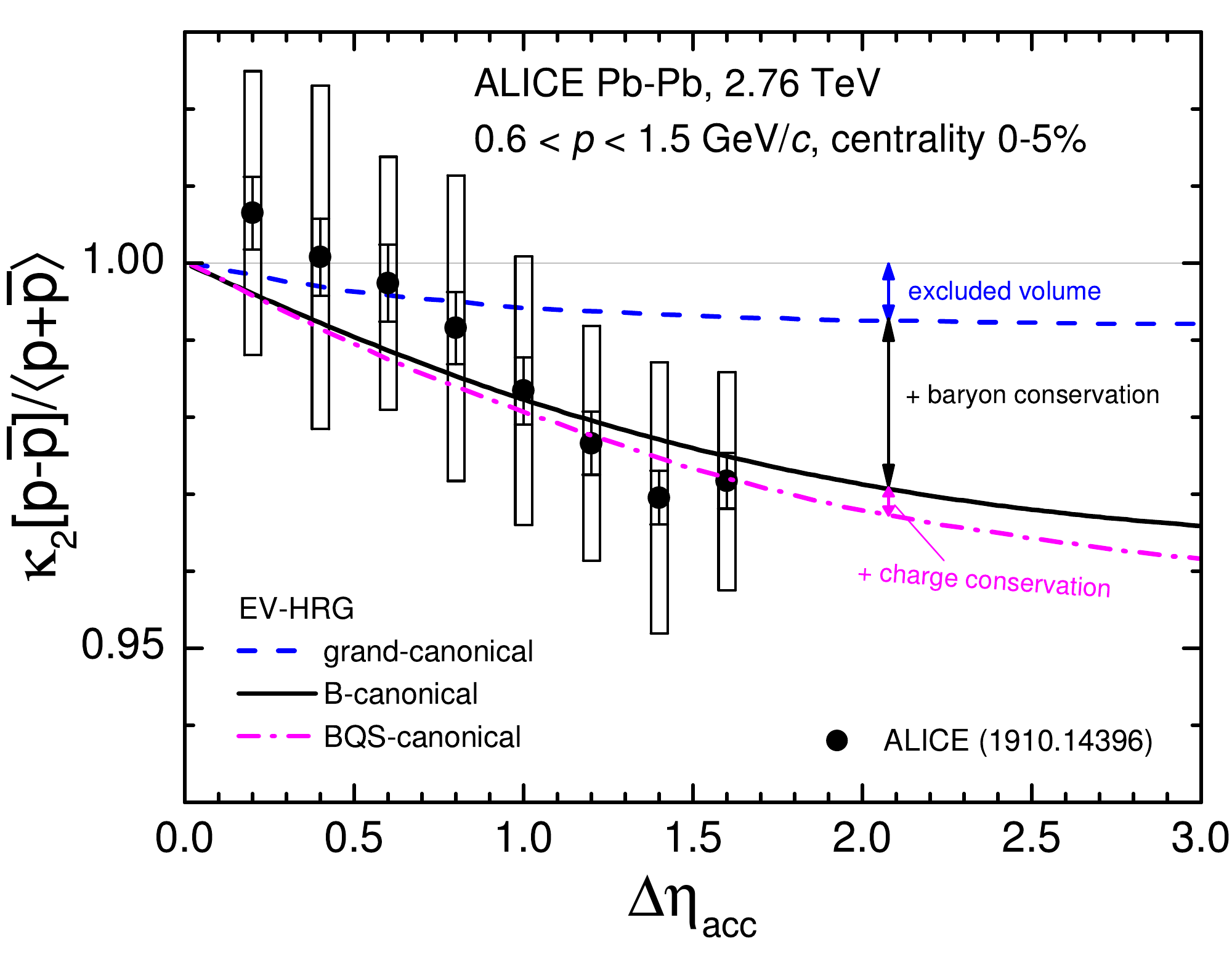}
  \includegraphics[width=.49\textwidth]{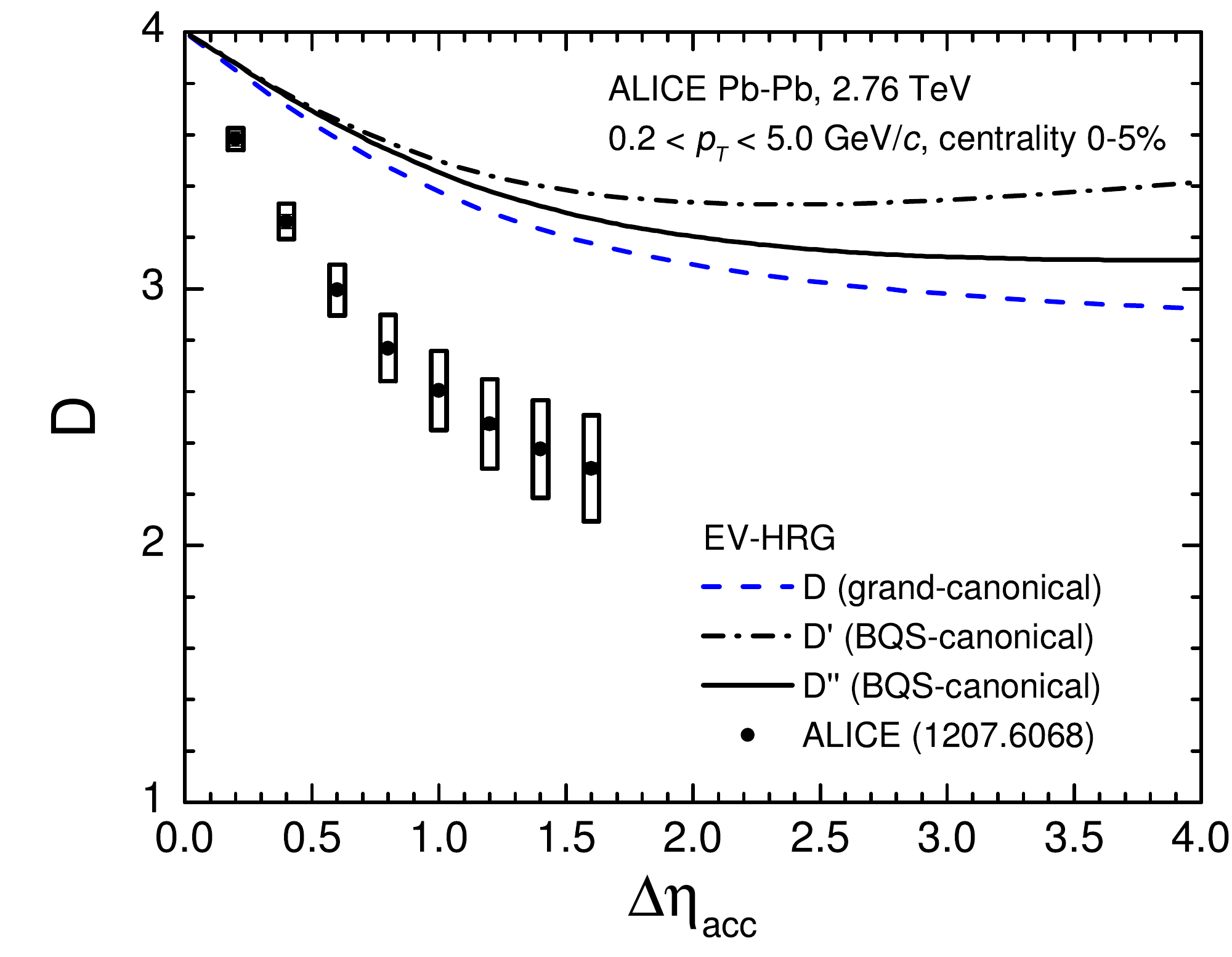}
  \caption{
  Pseudorapidity acceptance dependence of net proton $\kappa_2/\mean{p + \bar{p}}$~(\emph{left panel}) and the $D$-measure of net-charge fluctuations~(\emph{right panel}) in 0-5\% central Pb-Pb collisions at the LHC. 
  In the left panel the dashed blue line, the solid black line, and the dash-dotted magenta line corresponds to calculations within the grand-canonical, $B$-canonical, and $BQS$-canonical ensembles, respectively.
  In the right panel the dashed blue line and the black lines correspond to calculations within the grand-canonical, $BQS$-canonical ensembles, respectively.
  The $BQS$-canonical calculations are corrected for charge conservation via additive~(dash-dotted line) or multiplicative corrections~(solid line) corrections.
  The experimental data of the ALICE collaboration~\cite{Acharya:2019izy,Abelev:2012pv} are shown by the symbols with error bars.
  Adapted from Ref.~\cite{Vovchenko:2020kwg}.
  }
  \label{fig:ALICE}
\end{figure}

The results we have discussed so far correspond to fluctuations of baryons and protons in acceptances integrated over all transverse momenta.
This has not yet been achieved experimentally.
Instead, the ALICE collaboration has published measurements of the variance of the net-proton distribution in Pb-Pb collisions at $\sqrt{s_{\rm NN}} = 2.76$~TeV in an acceptance in a 3-momentum range $0.6 < p < 1.5$~GeV/$c$ and various longitudinal pseudorapidity ranges up to $|\eta| < 0.8$~\cite{Acharya:2019izy}.
The left panel of Fig.~\ref{fig:ALICE} depicts the comparison between the data~(symbols) and the EV-HRG model with exact baryon number conservation~(black line) for the ratio $\kappa_2/\mean{p+\bar{p}}$ of net protons in the ALICE acceptance.
In addition to the $B$-canonical calculation, we also depict the grand-canonical~(dashed blue line) and the full $BQS$-canonical~(dash-dotted magenta line) calculations.
This allows to elaborate on the influence of various effects on net proton fluctuations.
The results indicate that baryon number conservation has the strongest effect on net proton $\kappa_2/\mean{p+\bar{p}}$, followed by an additional suppression due to excluded volume interactions and exact conservation of electric charge.
When baryon conservation is included, the data are described within errors.
The uncertainties in the presently available measurements, as well as the limited momentum coverage, do not currently allow to distinguish any additional effects that go beyond baryon number conservation.
Fluctuation measurements of other identified hadron yields, such as $\Lambda$'s, kaons and pions, have also been performed by ALICE and the final results are being finalized~(see e.g. ~\cite{Arslandok:2020mda} for the preliminary results).
The influence of the various conservation laws, as well as resonance decays, on these observables have been discussed in Ref.~\cite{Vovchenko:2020kwg}.

Finally, we conclude the discussion of the experimental measurements at the LHC with the variance of net-charge distribution.
In Ref.~\cite{Abelev:2012pv} the ALICE collaboration has reported measurements of the so-called $D$-measure, which at the LHC is defined as
\eq{
D = 4 \frac{\mean{\delta Q^2}}{\mean{N_{\rm ch}}}~.
}
The measurements were performed in a broad transverse momentum range $0.2 < p_T < 5.0$~GeV/$c$ and varied pseudorapidity ranges.
As this quantity is affected by net-charge conservation, the measurements were corrected using either an additive correction~($D'$) of Ref.~\cite{Pruneau:2002yf} or a multiplicative correction~($D''$) advocated in~\cite{Bleicher:2000ek}, the difference between the two contributing to the systematic uncertainty.

The right panel of Fig.~\ref{fig:ALICE} depicts the pseudorapidity acceptance dependence of the $D$-measure resulting from Monte Carlo simulations using the subensemble sampler.
The grand-canonical calculation is depicted by the dashed blue line.
For the $BQS$-canonical calculation the two charge conservation corrections have been performed in the same way as it was done in the experiment, the results for $D'$ and $D''$ are shown by the black lines.
It is seen that the $BQS$-canonical $D''$ measure is closer to the true grand-canonical $D$ measure than $D'$, indicating that the multiplicative correction is more accurate.
The model, however, fails to describe the experimental data, which lies considerably below the model predictions.
The measurement, therefore, points to the suppression of net-charge fluctuations in central heavy-ion collisions at the LHC relative to the hadronic scenario.
One tantalizing possibility here is the QGP formation, where a suppression of the $D$-measure is expected~\cite{Jeon:2000wg}.
We hope that future measurements and analyses will shed more light on whether the observation of a suppressed $D$-measure constitutes a QGP signature.

\section{Summary and outlook}

In summary we have presented the subensemble sampler which is a new Cooper-Frye particlization routine appropriate for event-by-event fluctuations in heavy-ion collisions.
It is designed to sample multiplicity distribution of an arbitrary interacting hadron resonance gas, incorporating exact global conservation of the QCD conserved charges.
This allows to quantify the effects of global conservation, thermal smearing, and resonance decays on various cumulants measured in experiment, being an important step toward direct comparison between lattice QCD susceptibilities and heavy-ion measurements.
The method has been applied to provide insight into the behavior of various cumulants in heavy-ion collisions at the LHC.
This has been achieved by approximating the particlization stage by a boost-invariant blast-wave model with a cut-off at large rapidities and using hadron resonance gas model with excluded volume interactions in the baryon sector to reproduce the lattice QCD susceptibilities.

More data are available at lower collision energies, for instance from the RHIC beam energy scan program.
In particular, the higher-order cumulants of net-proton distribution have attracted a lot of attention, being potentially sensitive to the presence of the QCD critical point. 
The measurements by the STAR collaboration~\cite{Adam:2020unf,Abdallah:2021fzj} show significant  deviations of the cumulants from the uncorrelated proton production baseline. With the possible exception of the lowest energy, $\sqrt{s_{\rm NN}} = 7.7$~GeV, the results are qualitatively consistent with expectations from baryon number conservation. 
However, it should pointed out that presently no quantitative analysis of net-proton cumulants have been achieved in the hydrodynamic description of heavy-ion collisions in that energy range.
The particlization routine that we discussed is one way to achieve such a description.
Of course, at lower collision energies one can no longer assume the longitudinally boost-invariant blast-wave model to provide an appropriate quantitative description of the particlization surface.
On the other hand, recently three-dimensional hydro simulations of Au-Au collision at BES energies have become available, and they provide a reasonable description of the available rapidity distributions of various hadrons~\cite{Shen:2020jwv}.
The subensemble sampler can be combined with the numerical particlization hypersurfaces emerging from such hydro simulations to provide a quantitative analysis of net proton cumulants at RHIC BES.

\section*{Acknowledgements}
%\emph{Acknowledgments.} 
V.V. acknowledges the support by the
Feodor Lynen program of the Alexander von Humboldt
foundation.
This work received support through the U.S. Department of Energy, 
Office of Science, Office of Nuclear Physics, under contract number 
DE-AC02-05CH11231231 and within the framework of the
Beam Energy Scan Theory (BEST) Topical Collaboration.

\bibliography{proceed.bib}
% \printbibliography 

\end{document}